\documentclass[useAMS,usenatbib]{mn2e}

\usepackage{float}
\usepackage{graphicx}
\usepackage{grffile}
\usepackage{lineno}
\usepackage{multirow}
\usepackage{subfigure}
\usepackage{color}
\usepackage{cleveref}

\usepackage{amsmath,amssymb,latexsym}

\def\vFv{\nu F_{\nu}}
\def\Ep{E_{\rm p}}
\def\fermi{\it Fermi}

\def\fbb{F_{\rm BB}}
\def\ft{F_{\rm tot}}
\def\R{\fbb/\ft}

\title[Taking the Band Function Too Far]{ Taking the Band Function Too
  Far: A Tale of Two $\alpha$'s} \author[J. Michael
Burgess, Felix Ryde, and David Yu]{J. Michael Burgess$^{1,2}$\thanks{E-mail: jamesb@kth.se (JMB)}, Felix Ryde$^{1,2}$ and Hoi-Fung Yu$^{3,4}$\\
  $^{1}$The Oskar Klein Centre for Cosmoparticle Physics,
  AlbaNova, SE-106 91 Stockholm, Sweden\\
  $^{2}$Department of Physics, KTH Royal Institute of Technology,
  AlbaNova, SE-106 91 Stockholm, Sweden\\
  $^{3}$Max-Planck-Institut fur extraterrestrische Physik,
  Giessenbachstrasse 1, D-85748 Garching, Germany\\
  $^{4}$Excellence Cluster Universe, Technische Universit{\"a}t
  M{\"u}nchen, Boltzmannstra{\ss}e 2, 85748 Garching, Germany}
\begin{document}
\date{Accepted XXXX December XX. Received XXXX December XX; in original form XXXX October XX}

\pagerange{\pageref{firstpage}--\pageref{lastpage}} \pubyear{2014}

\maketitle

\label{firstpage}

\begin{abstract}

  The long standing problem of identifying the emission mechanism
  operating in gamma-ray bursts (GRBs) has produced a myriad of
  possible models that have the potential of explaining the
  observations.  Generally, the empirical Band function is fit to the
  observed gamma-ray data and the fit parameters are used to infer
  which radiative mechanisms are at work in GRB outflows. In
  particular, the distribution of the Band function's low-energy power
  law index, $\alpha$, has led to the so-called synchrotron
  ``line-of-death'' (LOD) which is a statement that the distribution
  cannot be explained by the simplest of synchrotron models alone.  As
  an alternatively fitting model, a combination of a blackbody in
  addition to the Band function is used, which in many cases provide a
  better or equally good fit. It has been suggested that such fits
  would be able to alleviate the LOD problem for synchrotron emission
  in GRBs.  However, these conclusions rely on the Band function's
  ability to fit a synchrotron spectrum within the observed energy
  band. In order to investigate if this is the case, we simulate
  synchrotron and synchrotron+blackbody spectra and fold them through
  the instrumental response of the $\fermi$ Gamma-ray Burst Monitor
  (GBM). We then perform a standard data analysis by fitting the
  simulated data with both Band and Band+blackbody models.  We find
  two important results: the synchrotron LOD is actually more severe
  than the original predictions: $\alpha_{\bf LOD} \sim
  -0.8$. Moreover, we find that intrinsic synchrotron+blackbody
  emission is insufficient to account for the entire observed $\alpha$
  distribution. This implies that some other emission mechanism(s) are
  required to explain a large fraction of observed GRBs.

\end{abstract}


\section[]{Introduction}

\begin{figure}
    \centering
    \includegraphics{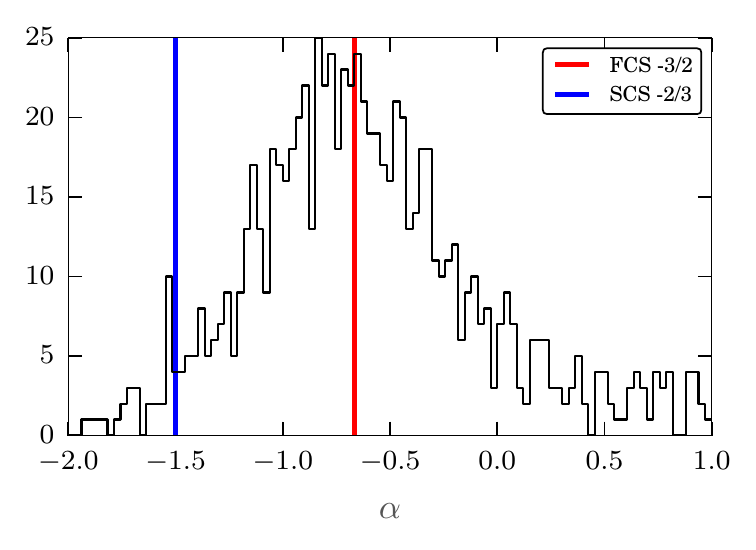}
    \caption{The GBM spectral catalog's $\alpha$ distribution
      \citep{Goldstein:2012} with the two standard LODs superimposed.}
    \label{fig:lod_cat}
  \end{figure}

While gamma-ray bursts (GRBs) are intrinsicly the brightest and most
energetic events in the Universe since the Big Bang, they are equally
one of the most ill understood. From energetics, the possible
progenitors, the collapse of supermassive population III stars or the
merger of two compact objects, can be heuristically argued for
\citep{Chevalier:1999,Ramirez-Ruiz:2002,Woosley:2006,Meszaros:2010},
but the pulse structure, observed spectra, and spectral evolution lack
a self-consistent theoretical explanation that can be bourne out by
the data \citep[e.g.][]{Preece:2014}. A key part of this problem is
the reliance on the fitted spectral parameters of the empirical Band
function \citep{Band:1993}, a smoothly broken power law used to fit
GRB spectral data, to infer the validity of their models. However, the
Band function lacks a physical origin and therefore deriving physical
implications from the fits relies on inferring what the various
spectral fit parameters are indicative of
\citep{Preece:1998,Ghirlanda:2003,Baring:2004,Daigne:2011}. In
general, the Band function can mimic several thermal and non-thermal
physical emissivities. The most commonly invoked example is relating
the Band function to the emission of optically-thin synchrotron by
relativistic electrons accelerated in the outflow of GRBs by magnetic
reconnection or shocks. The low-energy slope of synchrotron approaches
asymptotic values based on how fast the electrons are cooled by their
emission as they gyrate in a magnetic field. This can be separated
into two classes: fast-cooling (FCS) and slow-cooling (SCS)
synchrotron with low-energy photon number indices of $-3/2$ and $-2/3$
respectively \citep{Sari:1998}. With this consideration, the
distribution of low-energy slopes from GRB spectra, that have been fit
with the Band function, can be compared to the predicted low-energy
slopes and it can easily be seen that almost $1/3$ of all indices are
inconsistent with SCS and the nearly all are inconsistent with FCS
which has created the problem of the so-called ``lines-of-death''
(LOD) \citep{Crider:1998,Preece:1998,Kaneko:2006,Goldstein:2012} (see
Figure \ref{fig:lod_cat}).
  Such comparisons of Band's $\alpha$ index to various models have
  been a primary focus of modeling in the field. The GRB spectral
  catalog shows a peak in the $\alpha$ distribution of $\sim -1$ and
  many models try to achieve this central value
  \citep{Peer:2004,Peer:2006,Medvedev:2007,Beloborodov:2010,Daigne:2011,Uhm:2014}. Though,
  there is a substantial amount of spread in the $\alpha$ distribution
  and no one model has made predictions that can explain all observed
  values. Such predictions are essential if it is expected that there
  is a universal process that occurs in GRB jets. It may be that
  several types of emission processes are active and vary from burst
  to burst. However, the current lack of self-consistent simulations
  from progenitor to radiation production limit such a global
  assessment. Within the standard fireball model, it is very probable
  that several emission components can be present in the observed
  spectrum \citep{Meszaros:2000}, in particular emission from the
  photosphere and optically-thin regions could be superimposed upon
  one another.

  Recently, a trend has therefore evolved with the possibility of
  reconciling synchrotron emission with the Band $\alpha$ distribution
  which consists of fitting a blackbody (non-dissipative photosphere)
  in combination with the typically fitted Band function to spectra
  observed by GBM
  \citep{Guiriec:2011,Axelsson:2012,Iyyani:2013,Preece:2014}. This is
  a natural continuation of the fitting of a blackbody and a power law
  that occurred during the BATSE era
  \citep{Ryde:2004,Ryde:2005,Ryde:2009} however, with the expanded
  high-energy bandpass of GBM, the Band function and blackbody appear
  to be a more correct picture according to the data. The addition of
  the blackbody in some cases can change the $\alpha$ that was
  obtained by fitting the Band function alone to a value that is
  closer to what is expected from synchrotron. This is not, however, a
  universal observation \citep[for example, see the $\alpha$ values
  from ][]{Axelsson:2012}. Still, the changing of $\alpha$ values has
  the potential to alleviate the problem of the LOD by implying that
  the measured values of $\alpha$ from Band only fits are actually
  incorrect measurements and the spectral data should be fitted with a
  Band+blackbody model which will infer that the emission is actually
  a combination of synchrotron and a blackbody.

  There do exist predictions of emission coming from GRBs that have
  this non-thermal+blackbody spectrum which sufficiently motivates the
  fitting of Band+blackbody \citep{Meszaros:2000,Daigne:2002}. The
  significance of the observed blackbody has been calculated in many
  spectra \citep[e.g.][]{Axelsson:2012} and been shown to be quite
  high. It is also possible that the blackbody found in the spectra
  could arise from summing together the evolving spectrum of a single
  non-thermal emission mechanism. \citet{Burgess:2014c} showed that it
  is possible for an evolving Band function to introduce a blackbody
  into spectral fits if too long of a duration of the evolution is
  summed together in the fit and that time-resolved analysis is
  required to check for the existence of a blackbody in the spectral
  data. However, in any case, it is important ask what a spectrum that
  consists of either fast or slow-cooling synchrotron that has been
  folded through the GBM response looks like when fitted by the Band
  function.

  Herein, we investigate what the shape of the fitted Band function is
  when the intrinsic spectra consist of either fast or slow cooled
  synchrotron both with and without a blackbody by sythesizing these
  photon spectra and folding them through the GBM detector response
  and then fitting them with both Band and Band+blackbody photon
  functions. We are not primarily concerned about the quality of the
  fits but rather if the parameter distributions and values obtained
  from Band fits to actual physical photon models coincide with our
  assumptions. The article is divided as follows: in Section
  \ref{sec:lod} we simulate fast and slow-cooling synchrotron with
  $\vFv$ peaks sampled from the GBM peak flux catalog
  \citep{Goldstein:2012} to investigate the effect that the detector
  bandpass has on measuring the low-energy index of the synchrotron
  spectrum. In Section \ref{sec:bb} we simulate synchrotron spectra
  along with a blackbody where the synchrotron is held fixed and the
  blackbody flux is varied in $kT$ and flux to examine what the
  derived $\alpha$'s from Band fits would be under different
  scenarios.

  We stress that we are not addressing whether or not synchrotron and
  synchrotron+blackbody can arise in the GRB spectra from physical
  principles. Rather, we are testing the assertions that the parameter
  distributions from Band and Band+blackbody fits can be directly used
  to infer conclusions on the various underlying emission scenarios at
  work in GRB outflows.

\section[]{Testing the ``line-of-death''}
\label{sec:lod}

While the LOD is a strong motivator for model development in the field
of GRBs, it has not been tested directly by actually simulating what
the GRB spectral catalogs would look like if the spectra observed
actually came from either FCS or SCS emission. What we aim to test is
how the bandpass of the detector affects the measured Band $\alpha$
for the synchrotron models. Since the Band function's curvature around
the $\vFv$ peak differs from synchrotron and the synchrotron function
curves continuously below the $\vFv$ peak, then it is likely that
fitting a Band function to these physical spectra will have an effect
on $\alpha$ as the $\vFv$ peak approaches the low-energy edge of the
instrument's bandpass.

\begin{figure}
  \centering
  \includegraphics{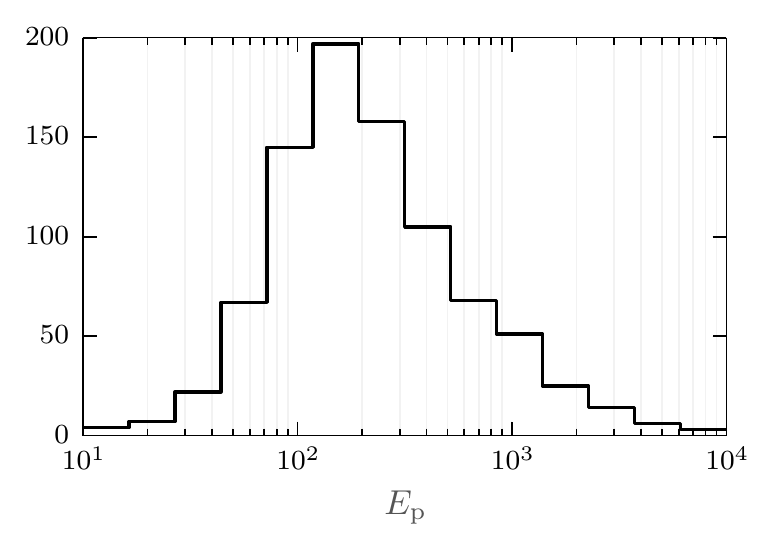}
  \caption{The GBM catalog $\Ep$ distribution.}
  \label{fig:EpCat}
\end{figure}

To examine this question, we sample the $\Ep$ distribution (see
Figure \ref{fig:EpCat}) of the $\fermi$ Gamma-ray Burst Monitor (GBM) peak
flux catalog \citep{Goldstein:2012} and use those values to simulate
synchrotron emission from fast- and slow-cooling electron
distributions (see Equations \ref{eq:slowE} and \ref{eq:fastE}). For
SCS, we assume that the electrons are distributed as a power law in
energy such that
\begin{equation}
  \label{eq:slowE}
  n_{\rm e}^{\rm slow}(\gamma)\propto \gamma^{-p}\;:\;\gamma_{\rm min}\leq\gamma
\end{equation}
where $p$ is the electron spectral index and $\gamma_{\rm min}$ is the
injection energy of the process that accelerates the electrons. The
spectrum of FCS arises when the electrons in the power law have cooled
quickly compared to the dynamical timescale via synchrotron emission
and pile up below the injection energy. This forms a broken power law
distribution of the electrons in energy of the form
\begin{equation}
  \label{eq:fastE}
  n_{\rm e}^{\rm fast}(\gamma) \propto\left\{%
\begin{array}{ll}
\gamma^{-2}&:\;\gamma_{\rm cool}<\gamma \leq \gamma_{\rm min}\\
\gamma^{1-p}&:\;\gamma<\gamma_{\rm min}
\end{array}
\right.
\end{equation}
where $\gamma_{\rm cool}$ is the energy to which the electrons cool
after a characteristic cooling time \citep[for a review on synchrotron
cooling see][]{Sari:1998,Burgess:2014a}. To compute the synchrotron
emission from these electron distributions, we convolve them with the
standard synchrotron kernel \citep{Blumenthal:1970}. For each sampled
$\Ep$ from the GBM catalog, we use the relation $\Ep\propto \Gamma B
\gamma_{\rm min}^2$, where $\Gamma$ and $B$ are the bulk Lorentz
factor and magnetic field strength respectively, to scale the $\vFv$
peak of the synchrotron spectrum. For the electron index, $p$ we
assume $p=3.5$ for slow-cooling and $p=2.5$ for fast-cooling to
recover the average observed value of the Band function's high-energy
index, $\beta\sim-2.2$.

With the derived photon spectra, we use detector responses from GBM to
produce count spectra for two Sodium-Iodide (NaI) and one
Bismuth-Germanante (BGO) detectors. Each simulated source spectrum has
a synthetic background added such that the signal-to-noise ratio is
30. The photon distribution of the background spectrum is a decreasing
power law in energy. In total, 1000 spectra are created for each of
the SCS and FCS models and then they are fit with the Band function
and their parameters recorded.

\begin{figure}
  \centering
\includegraphics{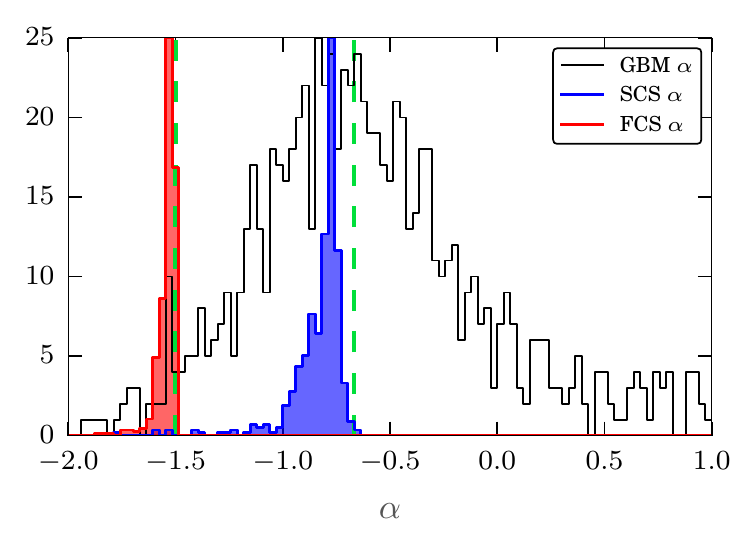}
  
\caption{The $\alpha$ distributions of the GBM peak flux spectral
  catalog with the $\alpha$ distributions from fast- and slow-cooling
  synchrotron superimposed. The \emph{green} lines indicate the LODs.}
\label{fig:lod_test}
\end{figure}

\begin{figure*}
  \centering  \subfigure[]{\includegraphics{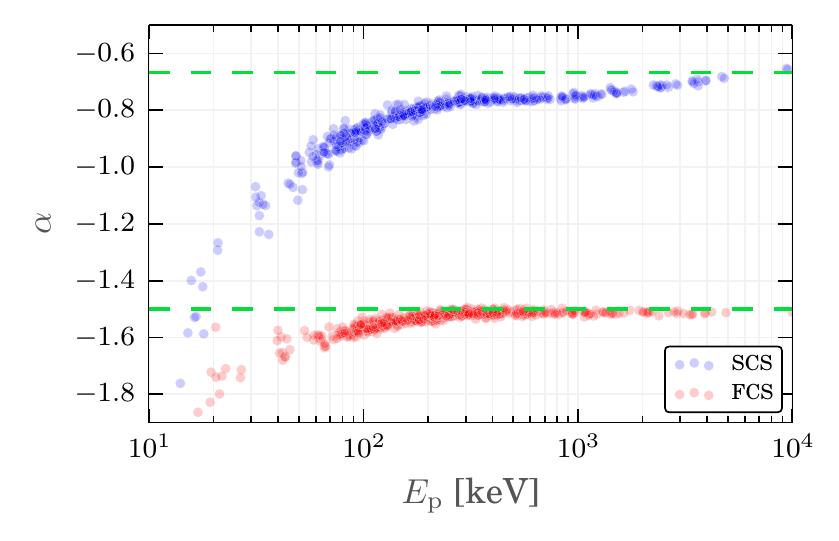}}\subfigure[]{\includegraphics{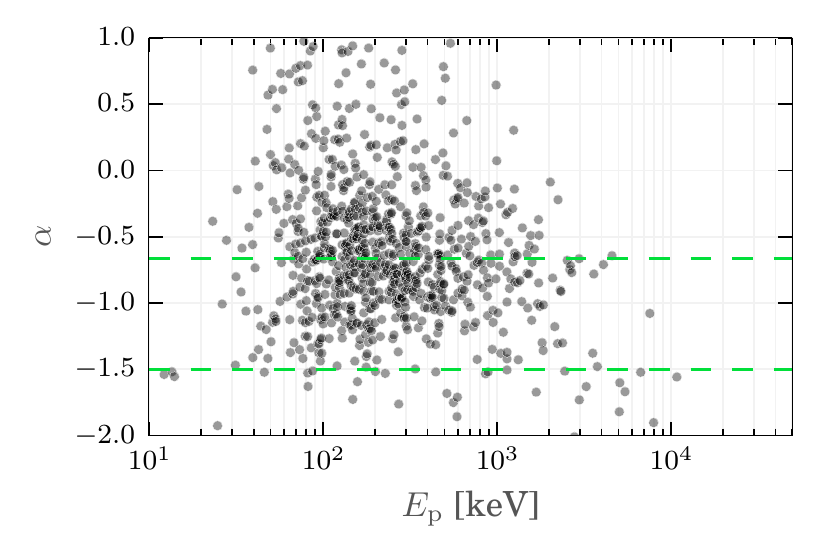}}
  \caption{$\Ep$ and $\alpha$ correlation for the FCS and SCS
    simulations. The \emph{left} panel demonstrates how the FCS
    (\emph{red}) and SCS (\emph{blue}) models would have $\alpha$
    measured if they were the intrinsic spectra. The \emph{right}
    panel is the actual GBM spectral catalog $\Ep$-$\alpha$
    correlation. In both panels, the LODs are shown by the
    \emph{green} lines.}
  \label{fig:EpAlpha}
\end{figure*}

Examining the distribution of $\alpha$'s found by simulating SCS, it
is clear that only a small portion of GBM spectra can be explained by
the model. We note that the LOD is actually \emph{worse} than what was
derived in \citep{Crider:1998,Preece:1998} because the SCS $\alpha$
distribution peaks at $\sim -.8$ as shown in Figure \ref{fig:lod_test}. This
leaves the harder half of the GBM distribution unreachable. The tail
of the distribution stretches towards more negative values. The spread
in the values of $\alpha$ derived from the Band fits to the SCS and
FCS synthetic spectra is attributed to the different values of $\Ep$
alone as can be seen in Figure \ref{fig:EpAlpha}. For the distribution of
$\alpha$'s from FCS, the LOD at $-3/2$ holds true and the width of the
distribution is narrow and stretches towards negative values. This is
because the FCS is very broad and its asymptotic power law behavior is
well approximated by a Band fit at low energies. Clearly, the two
standard synchrotron emission scenarios cannot account for the GBM
spectral catalog's $\alpha$ distribution.

%
%

\citet{Preece:1998} assumed the low-energy data was poorly described
by $\alpha$ because the Band function did not always approach their
asymptotic power law behavior if $\Ep$ was too close to the low-energy
bandpass of the detector. Therefore, they used the tangent slope of
the Band function at some fiducial value near the window to define an
effective power law slope ($\alpha_{\rm eff}$) that the authors
claimed was a better measure of the low-energy behavior of the
data. This is however, not what we are testing herein and the
correlation observed in Figure \ref{fig:EpAlpha} is due to spectral
curvature. We are testing how the Band function fit is affected by the
fact that synchrotron has a broader curvature than Band and this will
be sampled differently when the $\vFv$ peak of the spectrum is near
the low-energy bandpass of the detector.

\begin{figure}
  \centering
  \includegraphics{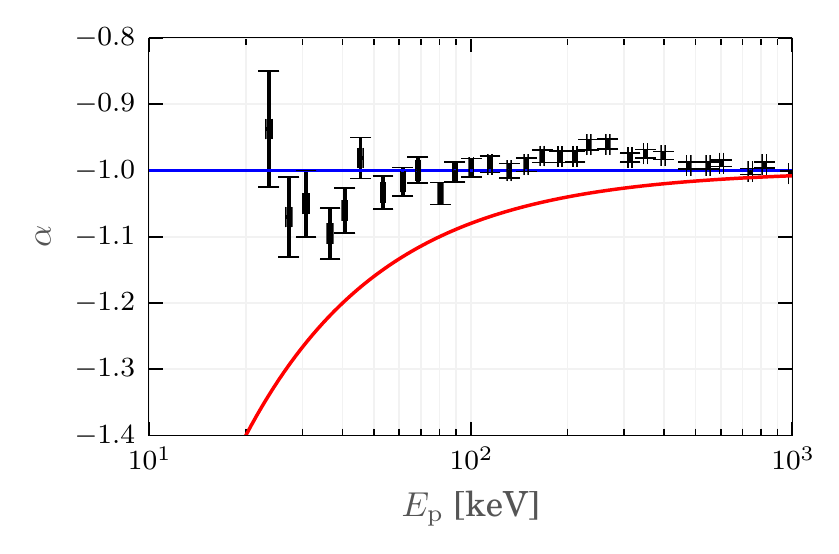}
  \caption{The effect of $\Ep$ being close to the low-energy bandpass
    on $\alpha$ is demonstrated. The simulated value of $\alpha$ is
    shown in \emph{blue} and the recovered values from Band fits are
    shown in \emph{black}. The $\alpha_{\rm eff}$ curve introduced in
    \citet{Preece:1998} is shown in \emph{red}. It clearly would
    artificially soften the correctly recovered $\alpha$.}
  \label{fig:aeff}
\end{figure}

Nevertheless, the effect in \citet{Preece:1998} could play an
important role in determining the low-energy behavior of the data. We
therefore test the ability of Band function fits to measure this
behavior. We simulate a Band function with $\Ep \in \left\{20;
  1000\right\}$ keV with $\alpha=-1$ and $\beta=-2.2$ and then fit
them with the Band function. We find that the $\alpha$ of the data is
recovered from the data regardless of the value of $\Ep$ (see
Figure \ref{fig:aeff}). Moreover, were we to use $\alpha_{\rm eff}$, we would
artificially soften the spectrum. Therefore, we concluded that the
Band function's natural $\alpha$ value is a appropriate to use for our
purposes.

\section[]{Can a blackbody fix the ``line-of-death''?}
\label{sec:bb}

The spectrum of a blackbody is uniquely set apart among astrophysical
emissivities by having a hard low-energy slope and the narrowest
$\vFv$ peak. Regardless of the physical implications of having
emission from GRBs in the form of synchrotron+blackbody, the addition
of a blackbody below the $\vFv$ peak of synchrotron affords the
opportunity to explain the harder values of $\alpha$ in the GBM
spectral catalog. There are reasons to take caution with using the
value of $\alpha$ to infer a emission
mechanism. \citet{Burgess:2014a}, for example, showed that even if
$\alpha$ from a Band fit to real GRB data has a value that corresponds
to FCS emission, a FCS photon model cannot fit the data because the
data around the $\vFv$ peak are too narrow for the broad curvature of
FCS. The point being that the curvature of the spectrum is as
important as the values of its asymptotic power law indices.
\begin{figure}
  \centering
  \includegraphics{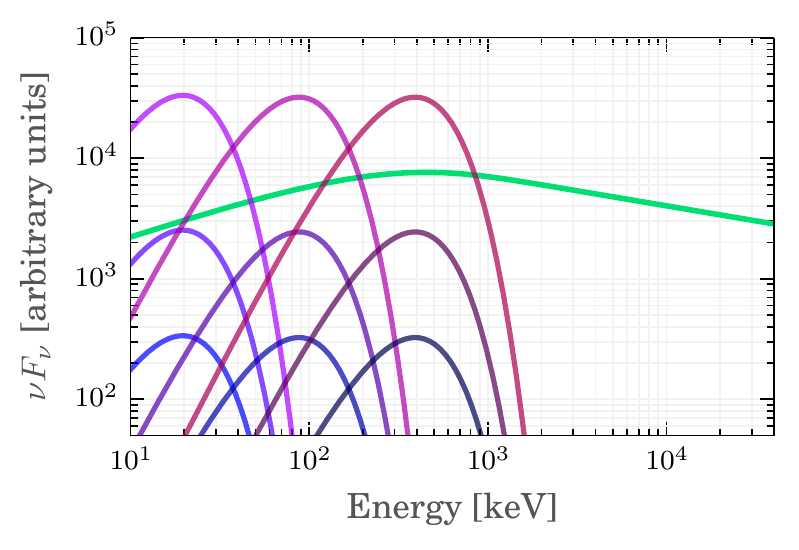}
  \caption{Demonstrating the grid of synthesized blackbodies (\emph{blue to purple}) with FCS (\emph{green}) superimposed.}
  \label{fig:grid}
\end{figure}

With curvature in mind, it is important to assess not only how adding
a blackbody component to the synchrotron would affect the $\alpha$ of
a Band fit to both components combined, but also whether or not the
combined curvature of the two components can be fit by the much
narrower curvature of the Band function. This is essential to
understanding if the model of synchrotron+blackbody can account for
the observed spectra. To investigate this problem, we simulate both
FCS and SCS held at a constant $\Ep$ and then add on a blackbody in a
grid of the blackbody temperature, $kT$ and the ratio of blackbody
energy flux ($\fbb$) to total energy flux ($\ft$) defined such that
when $R\equiv\R=1$, the blackbody accounts for the entire flux of the
spectrum. The grid of both $kT\in\{5;100\}$ keV and $R\in\{0.01;0.5\}$
(see Figure \ref{fig:grid}) span ranges that more than cover what has been
observed in the data
\citep{Guiriec:2011,Axelsson:2012,Iyyani:2013,Burgess:2014a}. For each
grid of blackbody parameters and each of the synchrotron models we
pick two values of fixed $\Ep$ for the synthesized synchrotron photon
spectra: $\Ep=300$ keV that represents the average observed value in
the data and $\Ep=1$ MeV to examine in greater detail what happens
below the $\vFv$ peak when a blackbody is added. In total, there are
four grids each with 900 synthetic spectra for all variations of the
blackbody parameters.

\subsection{Synthetic Slow-cooling Synchrotron + Blackbody}

\begin{figure*}
  \centering
  \subfigure[]{\includegraphics{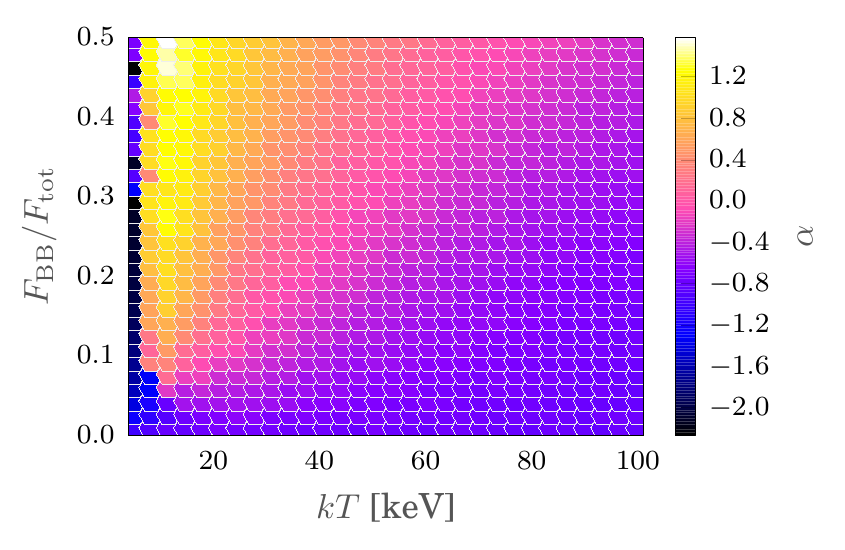}}\subfigure[]{\includegraphics{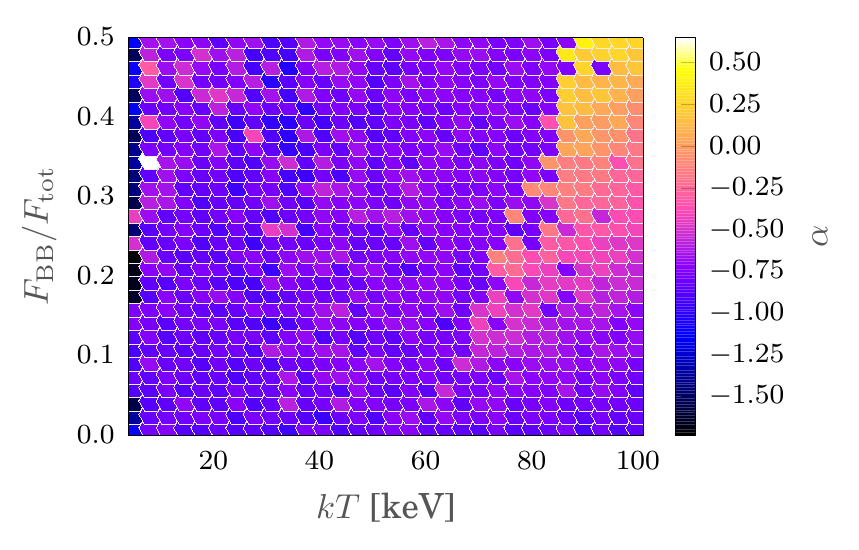}}
  \caption{The $\alpha$ distribtuions from the fits of Band
    (\emph{left}) and Band+blackbody (\emph{right}) to SCS+blackbody
    simulations with $\Ep=300$ keV.}
  \label{fig:scsbb300alpha}
\end{figure*}

\begin{figure*}
  \centering
  \subfigure[]{\includegraphics{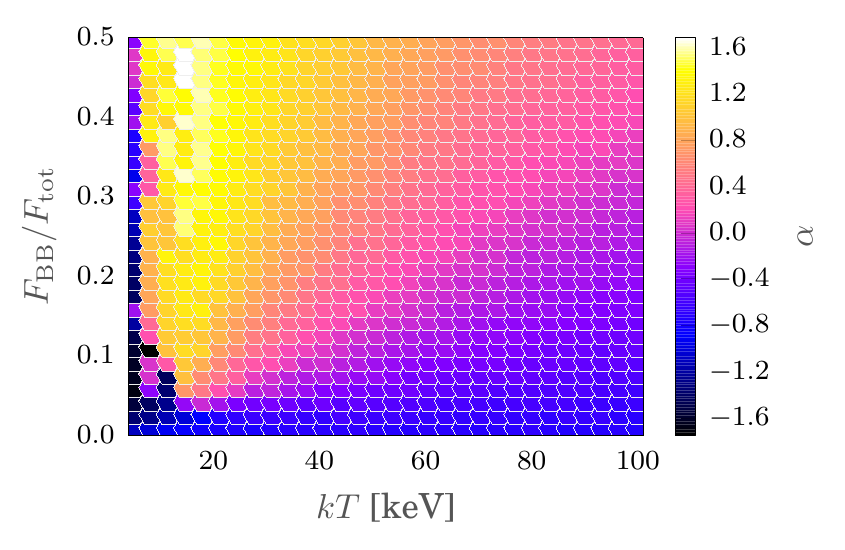}}\subfigure[]{\includegraphics{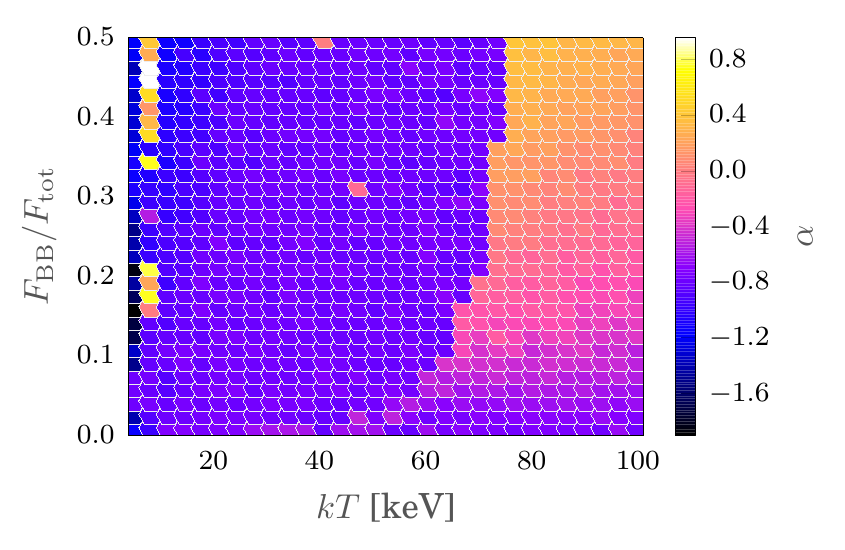}}
  \caption{Same as Figure \ref{fig:scsbb300alpha} but with $\Ep=1$ MeV.}
  \label{fig:scsbb1000alpha}
\end{figure*}

Due to its prolific use as an inference parameter for emission models,
we first examine the value of $\alpha$ obtained from Band only fits to
the synchrotron+blackbody simulations. Figure \ref{fig:scsbb300alpha}a and
Figure \ref{fig:scsbb1000alpha}a show the behavior of $\alpha$ as a function
of the blackbody parameters ($R$, $kT$) for $\Ep=300$ keV and $1$ MeV
respectively. For this grid, it is obvious that obtaining values of
$\alpha\geq 0$ requires low temperature blackbodies that must account
for a substantial fraction of the total energy flux. The key change
between the two values of simulated synchrotron $\Ep$ is that the
harder values of $\alpha$ are achieved for lower $R$ and higher $kT$
when $\Ep$ is greater.

\begin{figure}
  \centering
  \includegraphics{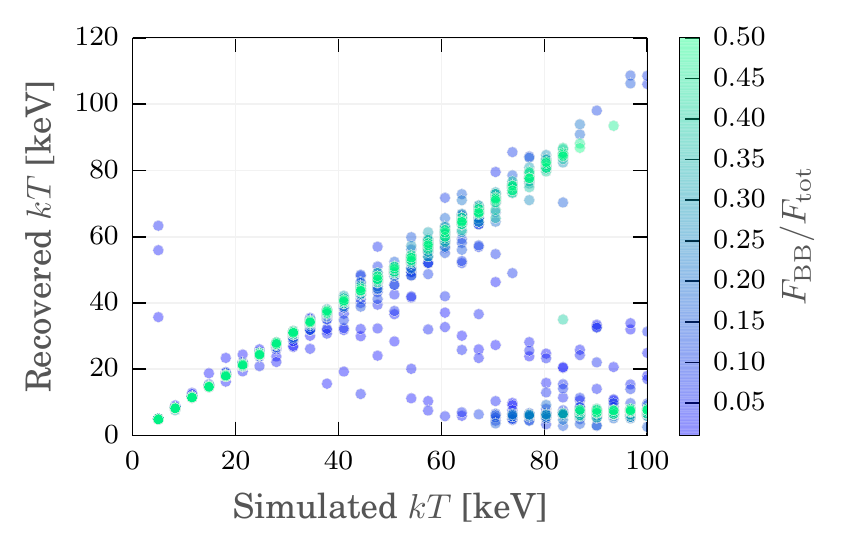}
  \caption{The simulated $kT$ of the blackbody vs. the $kT$ recovered when the spectrum is fitted by Band+blackbody when the simulated non-thermal spectrum is SCS with an $\Ep=300$ keV.}
  \label{fig:simKT}
\end{figure}

\begin{figure}
  \centering
  \includegraphics{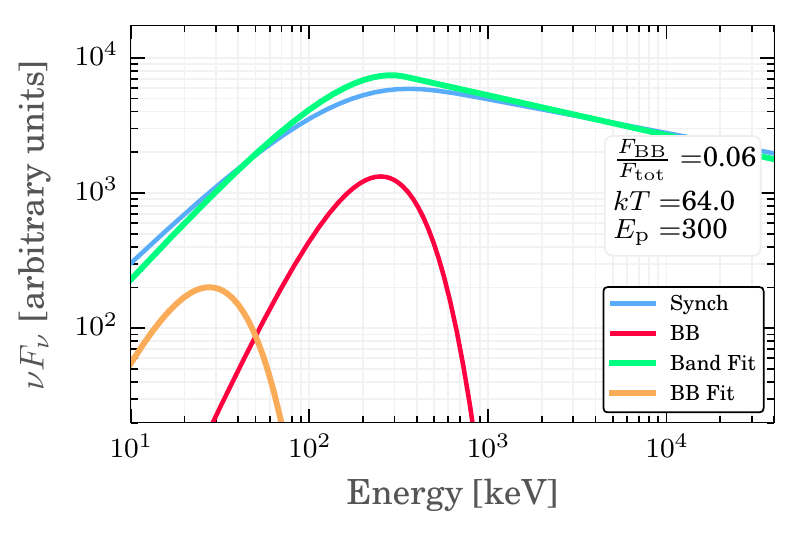}
  \caption{An example spectrum demonstrating the difference in the
    simulated blackbody and the one recovered in a Band+blackbody fit
    when the blackbody $\vFv$ peak coincides with the synchrotron
    $\vFv$ peak.}
  \label{fig:examplektbad}
\end{figure}

Figure \ref{fig:scsbb300alpha}b and Figure \ref{fig:scsbb1000alpha}b demonstrate
the behavior of $\alpha$ when the Band+blackbody model is fit the
simulated spectra. As expected, the value of $\alpha$ shifts to a more
SCS-like (-2/3) value except for high values of $kT$ due to the fact that at
high $kT$, the blackbody $\vFv$ peak coincides with synchrotron
peak. This makes it very hard for the fitting engine to fit both
components at their simulated values and lowers the value of $kT$
while increasing the value of $\alpha$, i.e., if a \emph{true}
blackbody in the data has a temperature that causes its $\vFv$ peak to
coincide with the peak of synchrotron (or perhaps any other
non-thermal emissivity), it is unlikely that a Band+blackbody fit will
find values that are indicative of the actual physical spectrum (see
Figures \ref{fig:simKT} and \ref{fig:examplektbad}).

\begin{figure*}
  \centering
  \subfigure[]{\includegraphics{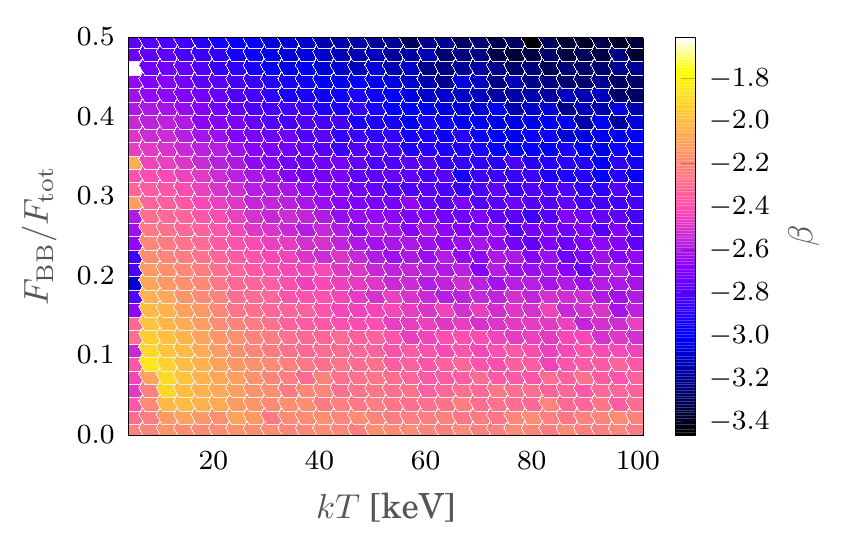}}\subfigure[]{\includegraphics{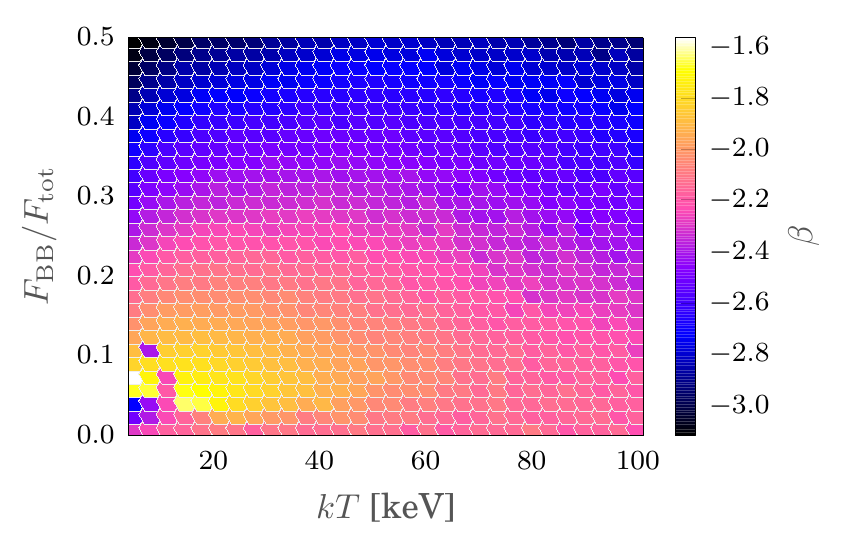}}
    \caption{The values of $\beta$ recovered for the different
      blackbody parameters with $\Ep=300$ keV (\emph{left}) and
      $\Ep=1$MeV (\emph{right}).}
  \label{fig:scsbeta}
\end{figure*}

\begin{figure}
  \centering
  \includegraphics{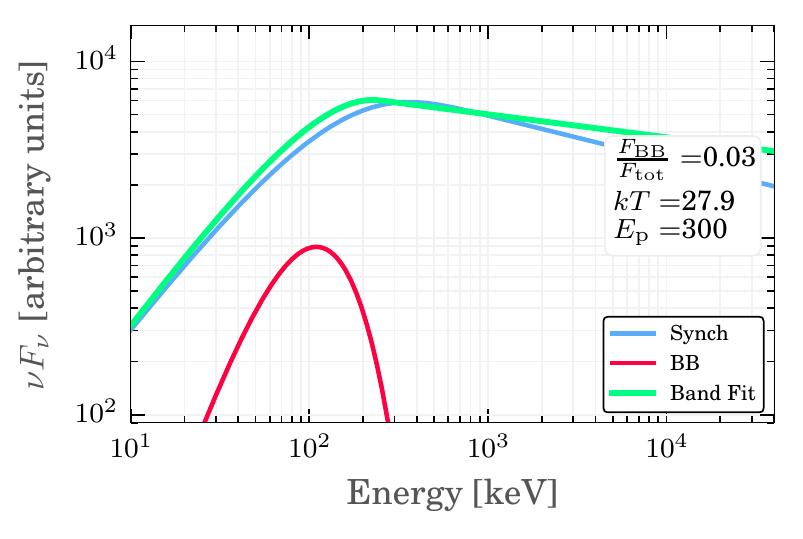}
  \caption{An example spectrum demonstrating the Band function's
    inability to properly fit the broad curvature of the simulated
    SCS+blackbody spectrum. The Band function adjusts by lowering
    $\Ep$ and increasing $\beta$ to values not typically found in the
    GBM catalog.}
\label{fig:exbetascs}
\end{figure}

\begin{figure}
  \centering
  \includegraphics{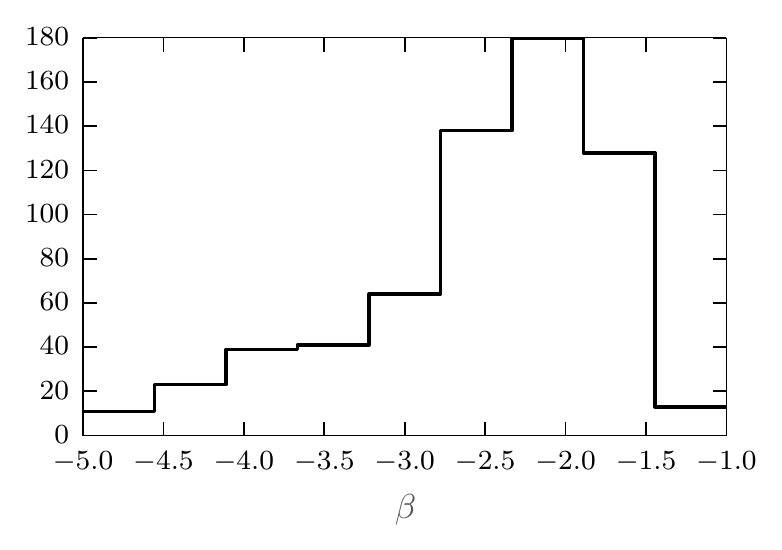}
  \caption{The GBM catalog $\beta$ distribution.}
  \label{fig:betaCat}
\end{figure}

To understand how the combined curvature of the simulated
SCS+blackbody spectrum affects the fit of Band only to the data, we
examine how the high-energy power law index of the Band function
($\beta$) is affected by the addition of the
blackbody. Figure \ref{fig:scsbeta} shows the recovered value of $\beta$ for
the different blackbody parameters. Very hard values ($\beta\geq-2.$)
are found for values of the blackbody typically observed in the data
($kT\simeq 30$ keV and $R\simeq 0.1$). This is due to the broad
curvature of the SCS+blackbody spectrum that cannot be fit with the
narrower Band function. To compensate, the Band function $\Ep$ is
lowered and $\beta$ increased (see Figure \ref{fig:exbetascs}). Such values
of $\beta$ are rare in the spectral catalog (see Figure \ref{fig:betaCat}).

\begin{figure*}
  \centering
  \subfigure[]{\includegraphics{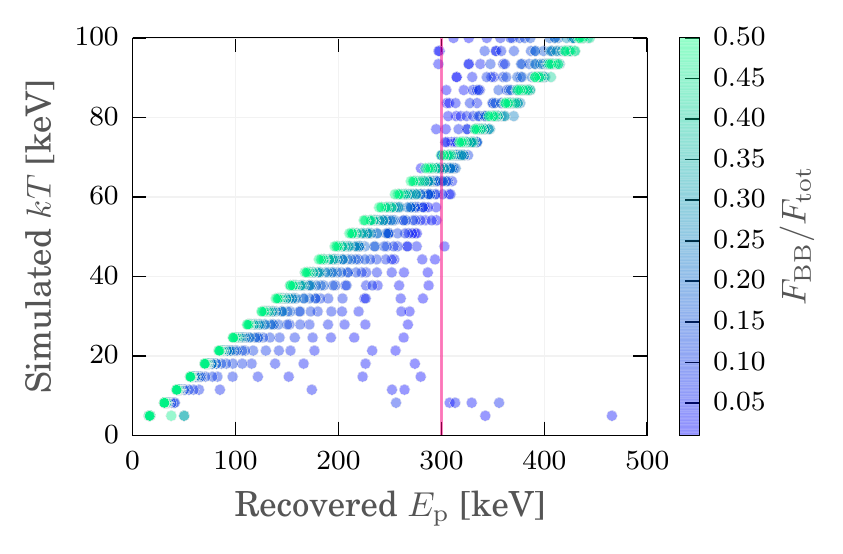}}\subfigure[]{\includegraphics{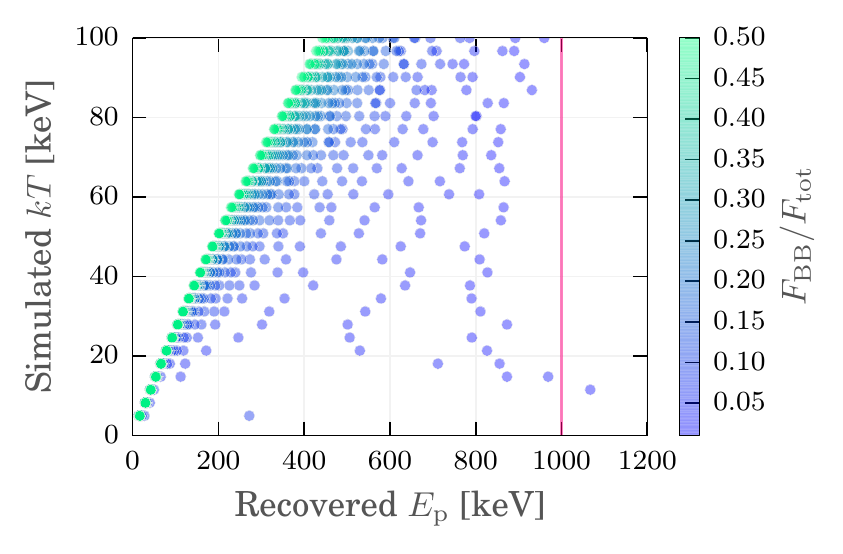}}
  \caption{The value of the recovered Band $\Ep$ as a function of the
    simulated blackbody $kT$ when the simulated SCS spectrum has an
    $\Ep=300$ keV (\emph{left}) and $\Ep=1$ MeV (\emph{right}). The
    \emph{pink} line indicates the simulated value of the synchrotron
    $\Ep$. }
  \label{fig:scsrecEp}
\end{figure*}

\begin{figure*}
  \centering
  \subfigure[]{\includegraphics{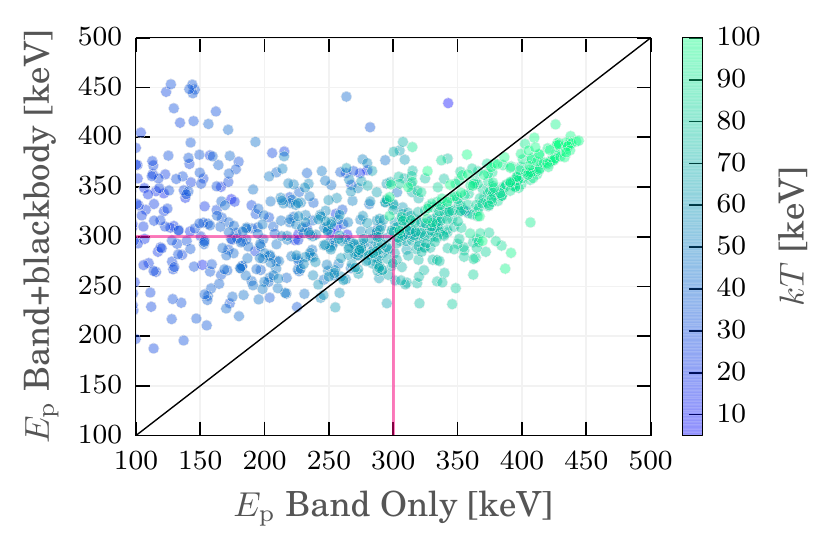}}\subfigure[]{\includegraphics{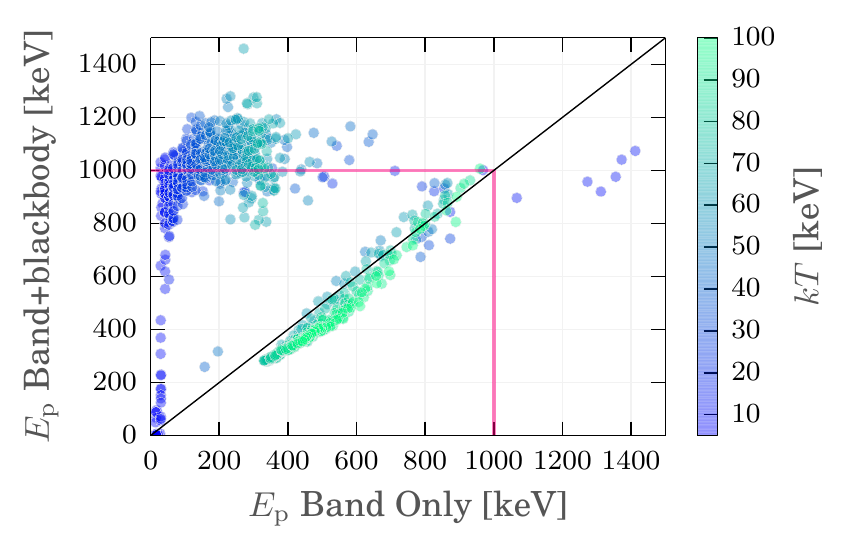}}
  \caption{The shift in recovered $\Ep$ as a function of $kT$ when
    fitting Band+blackbody to the SCS+blackbody simulations when the
    simulated SCS spectrum has an $\Ep=300$ keV (\emph{left}) and
    $\Ep=1$ MeV (\emph{right}). The \emph{pink} lines indicate the simulated
    value of the synchrotron $\Ep$. Points above the black line shift
    to higher $\Ep$ values when fit with Band+blackbody as opposed to
    Band.}
  \label{fig:epshiftscskt}
\end{figure*}

\begin{figure*}
  \centering
 
\subfigure[]{\includegraphics{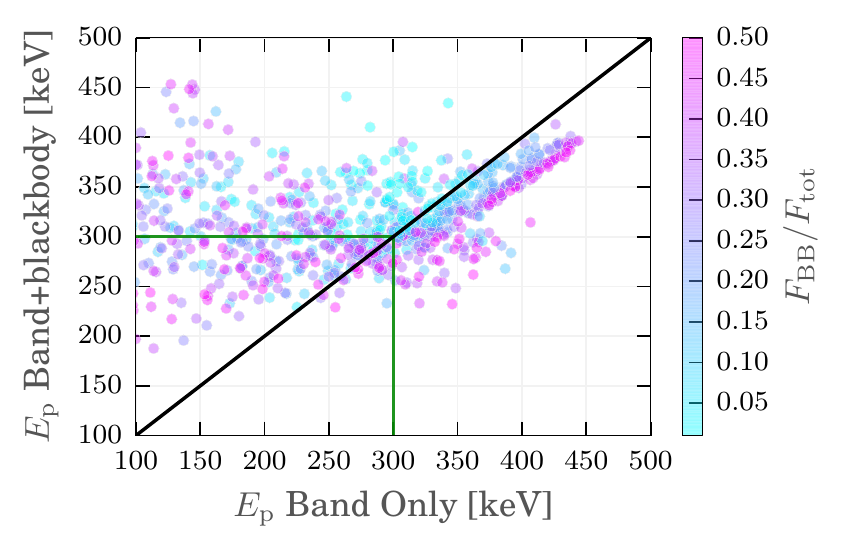}}\subfigure[]{\includegraphics{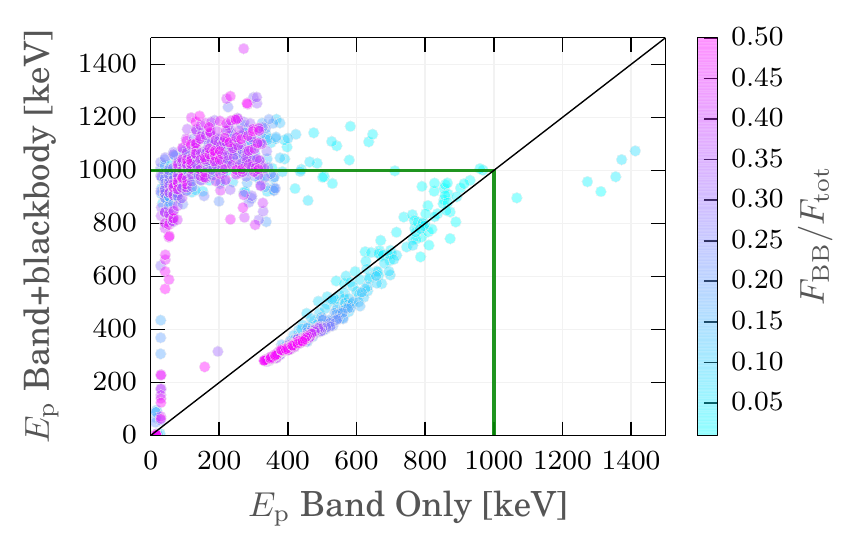}}
\caption{Same as Figure \ref{fig:epshiftscskt} but showing how the $\Ep$
  shift is affected by $R$. The \emph{green} lines indicate the simulated
  value of the synchrotron $\Ep$.}
  \label{fig:epshiftscsfrac}
\end{figure*}

Finally, the behavior of $\Ep$ is examined. When fitting Band only to
the synthetic spectra where the synchrotron $\Ep=300$ keV,
Figure \ref{fig:scsrecEp}a shows the Band $\Ep$ is sensitive to the
blackbody $kT$ except when $R$ is very low. However, when the
synchrotron $\Ep$ is increased to 1 MeV, Figure \ref{fig:scsrecEp}b shows
the recovered Band $\Ep$ to be sensitive to $kT$ nearly independent of
$R$. Next, the shift in the Band $\Ep$ when the Band+blackbody model
is fitted appears to be more correlated with $kT$ than $R$ (see
Figurea \ref{fig:epshiftscskt} and \ref{fig:epshiftscsfrac}). For lower values of $kT$
(similar to values recovered in the real observations) $\Ep$ shifts
systematically to lower values while for high values of $kT$, the
shift of $\Ep$ is to higher values. It is important to notice that for
both low and high $kT$ the value of $\Ep$ recovered by the
Band+blackbody fit is not always accurate and can \emph{vary greatly
  from the simulated values}.

\subsection{Synthetic Fast-cooling Synchrotron + Blackbody}

\begin{figure*}
  \centering
  \subfigure[]{\includegraphics{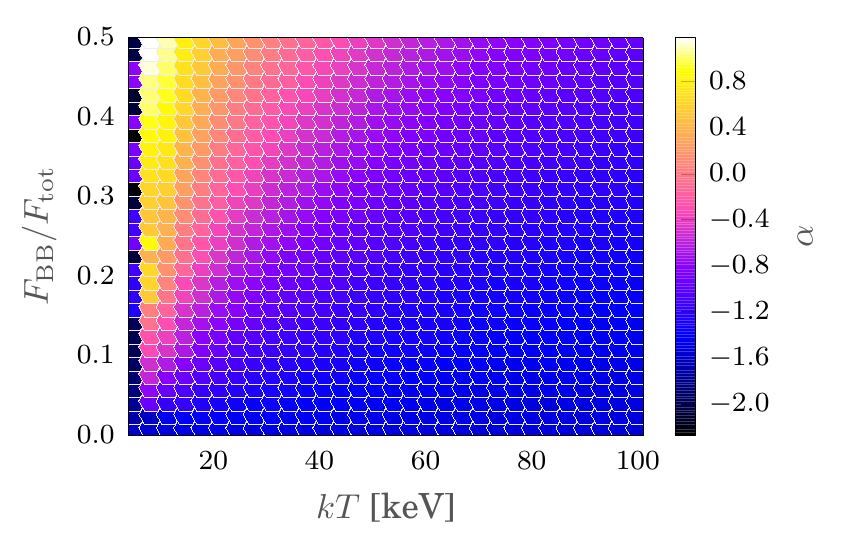}}\subfigure[]{\includegraphics{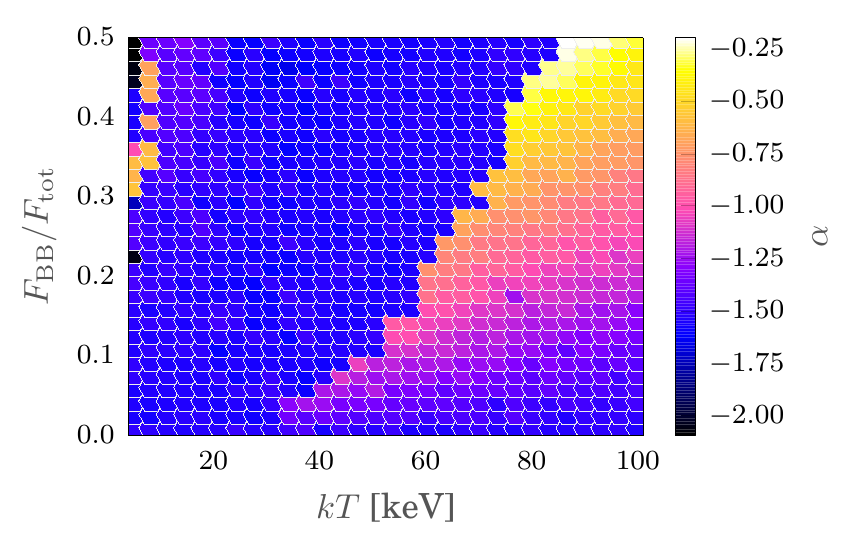}}
  \caption{The $\alpha$ distribtuions from the fits of Band
    (\emph{left}) and Band+blackbody (\emph{right}) to FCS+blackbody
    simulations with $\Ep=300$ keV.}
  \label{fig:fcsbb300alpha}
\end{figure*}

\begin{figure*}
  \centering
  \subfigure[]{\includegraphics{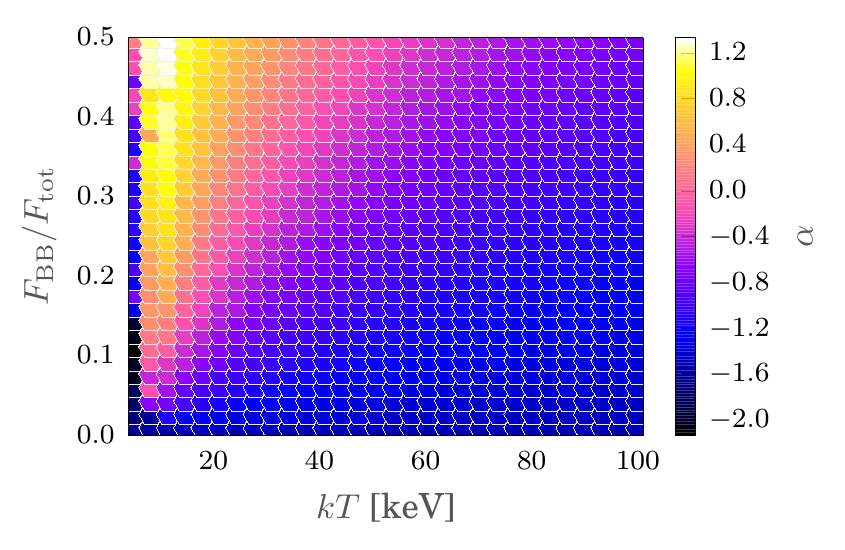}}\subfigure[]{\includegraphics{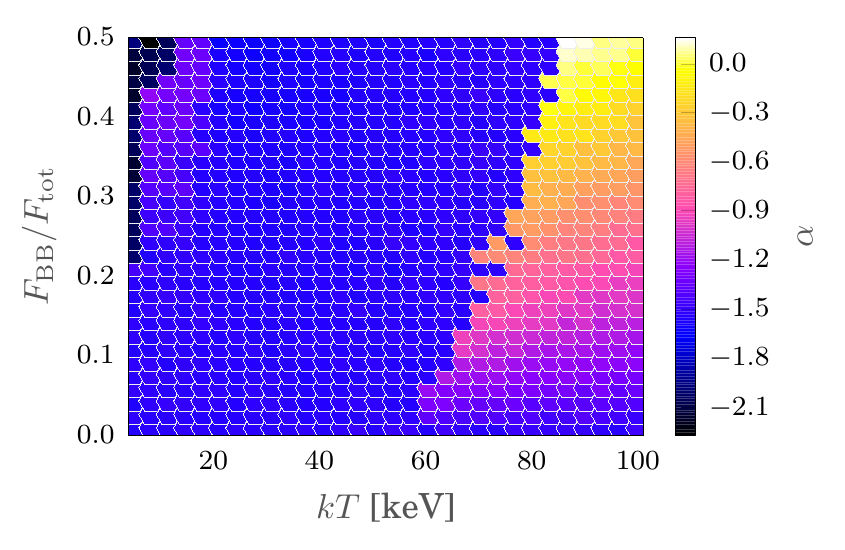}}
  \caption{Same as Figure \ref{fig:fcsbb300alpha} but with $\Ep=1$ MeV.}
  \label{fig:fcsbb1000alpha}
\end{figure*}

\begin{figure}
  \centering
  \includegraphics{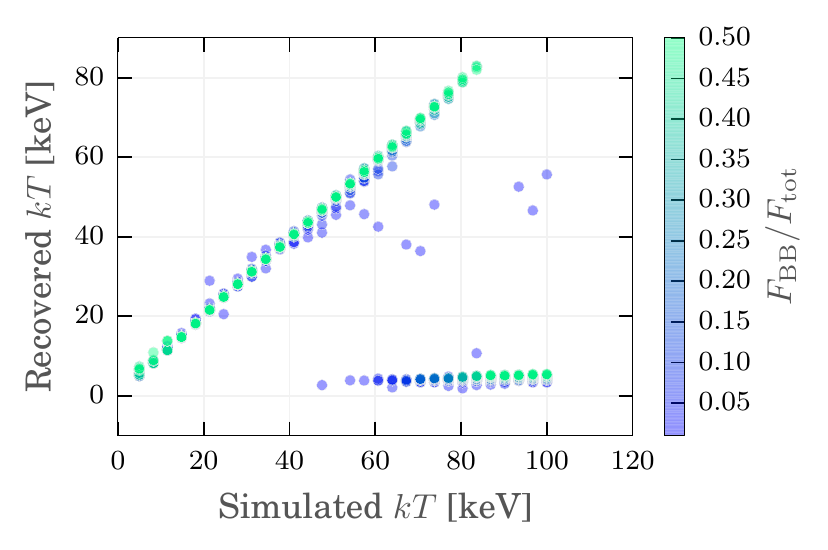}
  \caption{The simulated $kT$ of the blackbody vs. the $kT$ recovered when the spectrum is fitted by Band+blackbody when the simulated non-thermal spectrum is FCS with an $\Ep=1$ MeV.}
  \label{fig:simKTfast}
\end{figure}

\begin{figure}
  \centering
  \includegraphics{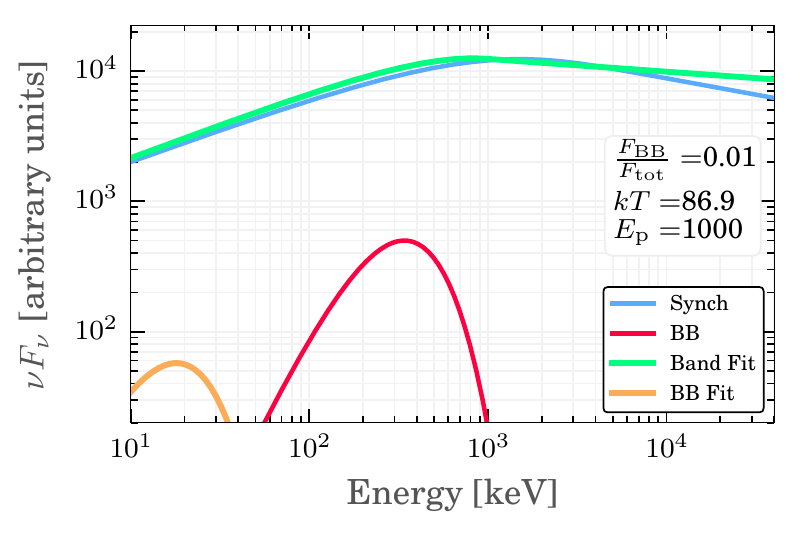}
  \caption{An example spectrum demonstrating the difference in the
    simulated blackbody and the one recovered in a Band+blackbody fit
    when the blackbody $\vFv$ peak coincides with the synchrotron
    $\vFv$ peak.}
  \label{fig:examplektbadfast}
\end{figure}

The FCS+blackbody simulations exhibit many of the same features found
for SCS with minor adjustments for the values of $\alpha$ found in the
Band only fits with respect to the simulated blackbody parameters (see
Figures \ref{fig:fcsbb300alpha} and \ref{fig:fcsbb1000alpha}). As with SCS, achieving
values of $\alpha >-0.8$ requires $kT \lesssim 60$ keV and $R\gtrsim
.1$, which does not coincide with observations of blackbodies in GRB
spectra. However, when the spectra are fit with the Band+blackbody
model, the measured $\alpha$ shifts to what is expected for FCS unless
the blackbody $\vFv$ peak coincides with the FCS $\Ep$. This can be
seen just as with SCS via Figure \ref{fig:simKTfast} and \ref{fig:examplektbadfast}.

\begin{figure*}
  \centering
  \subfigure[]{\includegraphics{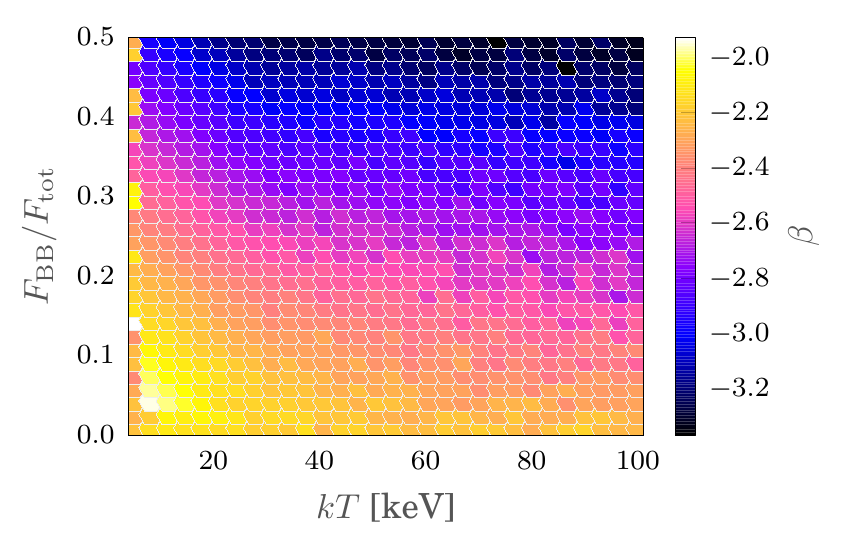}}\subfigure[]{\includegraphics{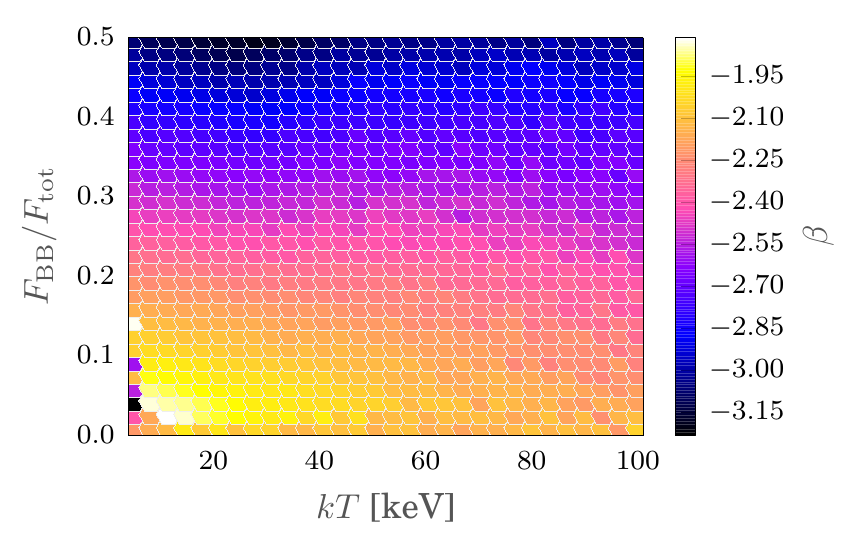}}
  \caption{The values of $\beta$ recovered for the different blackbody
    parameters with $\Ep=300$ keV (\emph{left}) and $\Ep=1$MeV
    (\emph{right}). The \emph{pink} line indicates the simulated value of the
    synchrotron $\Ep$.}
  \label{fig:fcsbeta}
\end{figure*}

\begin{figure}
  \centering
  \includegraphics{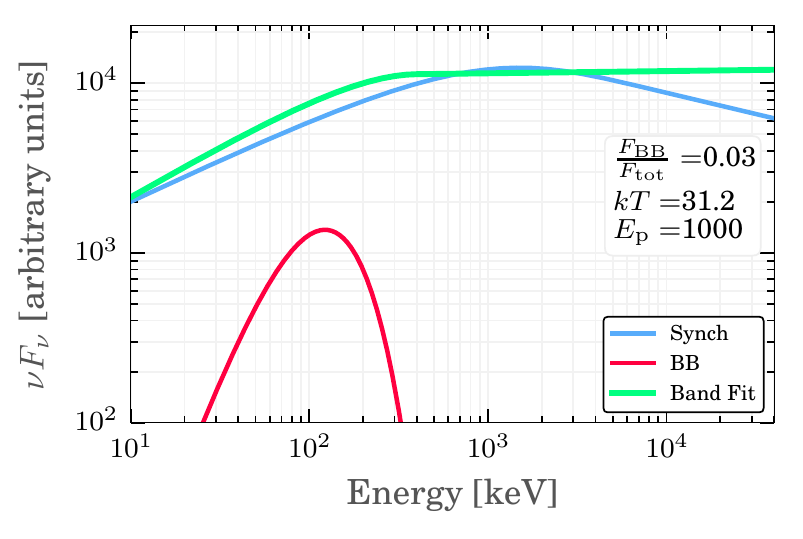}
  \caption{An example spectrum demonstrating the Band function's
    inability to properly fit the broad curvature of the simulated
    FCS+blackbody spectrum. The Band function adjust by lowering $\Ep$
    and increasing $\beta$ to values not typically found in the GBM
    catalog.}
\label{fig:exbetafcs}
\end{figure}

The behavior of the recovered value of $\beta$ from the simulations is
slightly altered from what is observed with SCS+blackbody as shown in
Figures \ref{fig:fcsbeta} and \ref{fig:exbetafcs}. Whereas high $kT$ and low $R$ result
in acceptable $\beta$ values for SCS+blackbody, the value of $\beta$
for FCS+blackbody is mostly sensitive to $R$. The already broad
curvature of FCS is not affected so much by where the blackbody $\vFv$
peak appears in energy as it is affected by the blackbody's
flux. Still, the curvature is too wide for the Band function to fit
the spectrum without increasing $\beta$ to typically unobserved,
higher values.


\begin{figure*}
  \centering
  \subfigure[]{\includegraphics{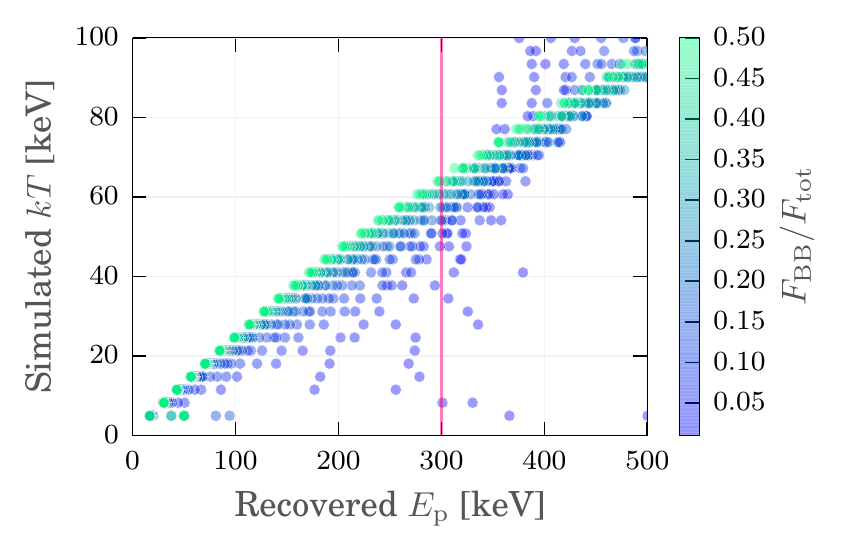}}\subfigure[]{\includegraphics{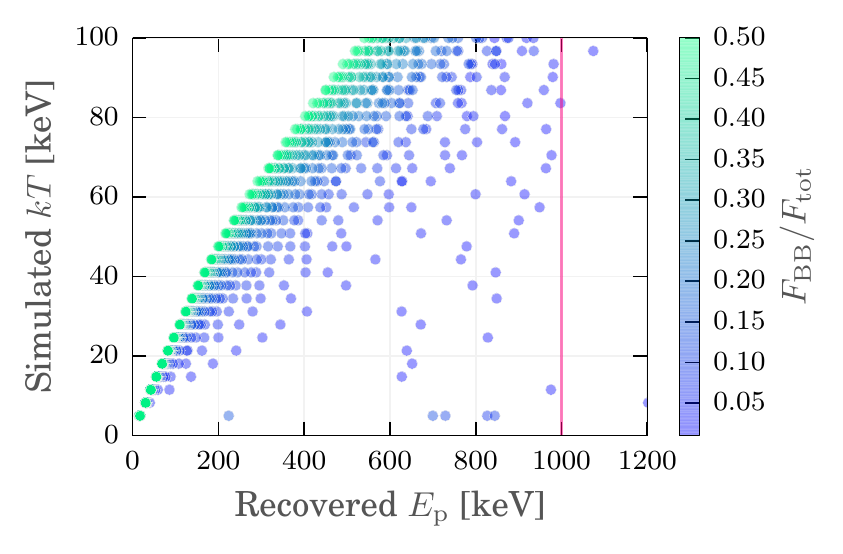}}
  \caption{The value of the recovered Band $\Ep$ as a function of the
    simulated blackbody $kT$ when the simulated FCS spectrum has an
    $\Ep=300$ keV (\emph{left}) and $\Ep=1$ MeV (\emph{right}). The
    \emph{pink} line indicates the simulated value of the synchrotron $\Ep$.}
  \label{fig:fcsrecEp}
\end{figure*}

\begin{figure*}
  \centering
  \subfigure[]{\includegraphics{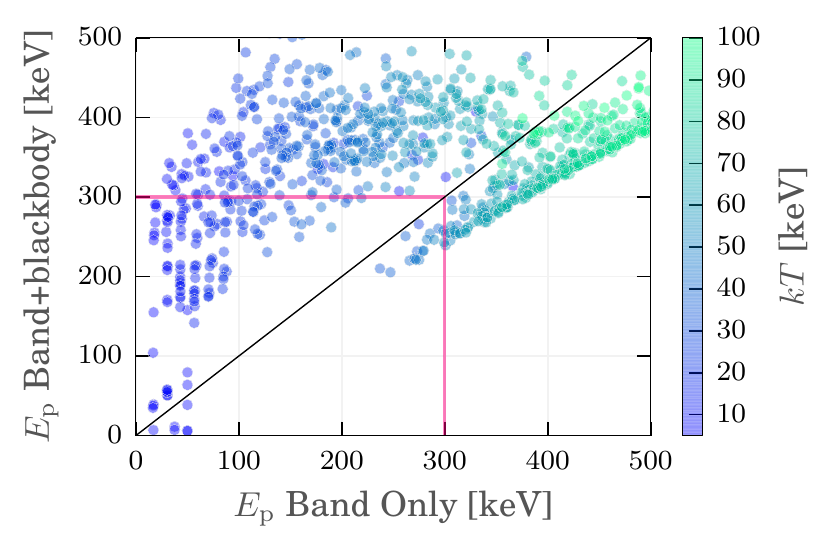}}{\includegraphics{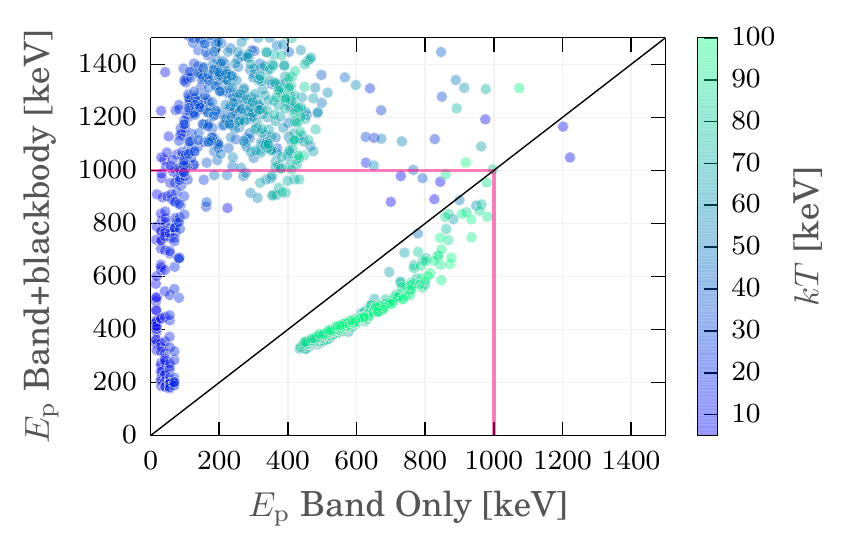}}
  \caption{The shift in recovered $\Ep$ as a function of $kT$ when
    fitting Band+blackbody to the FCS+blackbody simulations when the
    simulated SCS spectrum has an $\Ep=300$ keV (\emph{left}) and
    $\Ep=1$ MeV (\emph{right}). The \emph{pink} line indicates the simulated
    value of the synchrotron $\Ep$. Points above the black line shift
    to higher $\Ep$ values when fit with Band+blackbody as opposed to
    Band.}
  \label{fig:epshiftfcskt}
\end{figure*}

\begin{figure*}
  \centering
 
\subfigure[]{\includegraphics{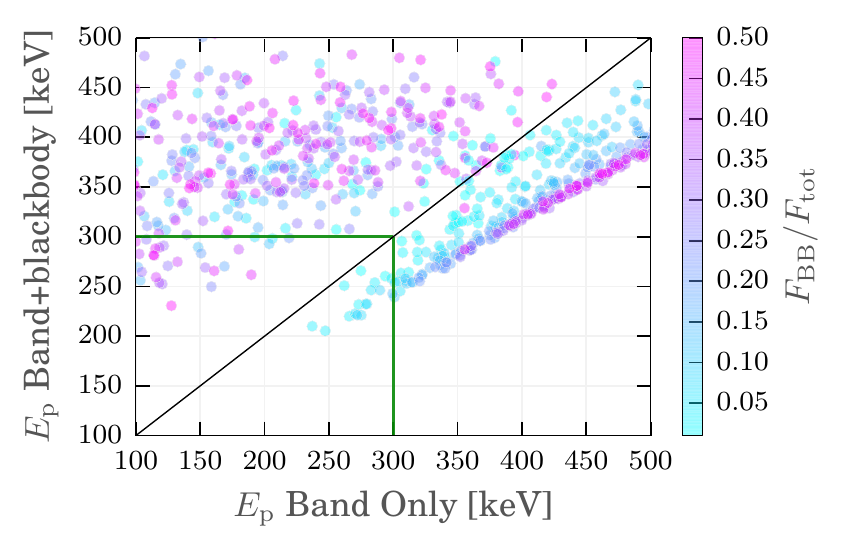}}\subfigure[]{\includegraphics{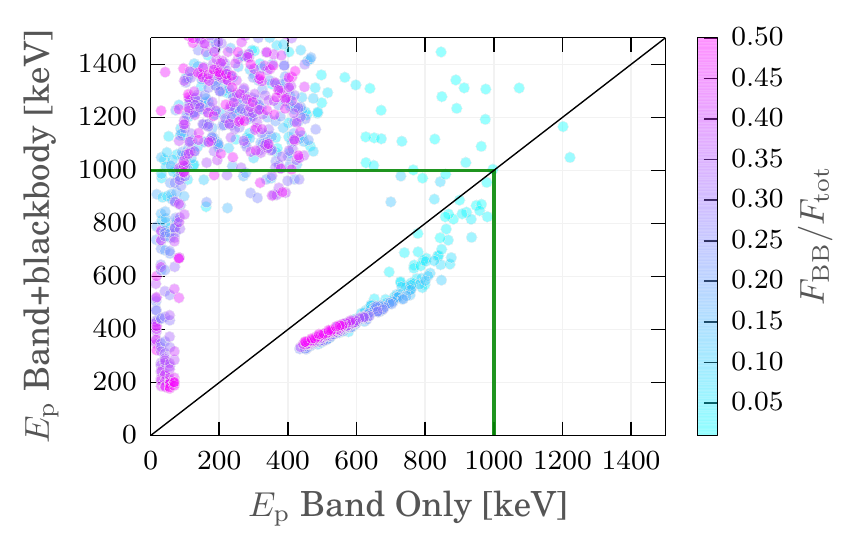}}
\caption{Same as Figure \ref{fig:epshiftfcskt} but showing how the $\Ep$
  shift is affected by $R$. The \emph{green
} line indicates the simulated
    value of the synchrotron $\Ep$.}
  \label{fig:epshiftfcsfrac}
\end{figure*}

The value of $\Ep$ from Band only fits is \emph{always} affected by
the $kT$ regardless of the flux of the blackbody as is shown in
Figure \ref{fig:fcsrecEp}. Again, this is an effect of the broad curvature
of the FCS spectrum. Figures \ref{fig:epshiftfcskt} and \ref{fig:epshiftfcsfrac} show a
similar behavior as with SCS for the shift in the recovered $\Ep$ when
fitting with Band or Band+blackbody but is slightly more sensitive to
the flux of the simulated blackbody though the overall dependence is
still from changing $kT$.

Overall, we find similar behaviors for the fits of Band and
Band+blackbody to the simulated SCS and FCS photons models. The main
change occurs in the values of $\alpha$ and $\beta$ recovered as a
function of the blackbody parameters. The relationships found in the
test all have similar dependencies with the magnitudes of the effect
changed due to the different curvature of the SCS and FCS spectra.


\section{Discussion}

Herein, we have investigated the ability of both optically-thin
synchrotron emission in the form of slow-cooling and fast-cooling as
well as synchrotron emission with an additional blackbody (photosphere)
to explain the observed Band parameter distributions in the GBM
spectral catalogs. Additionally, we have investigated in detail the
properties one would observe if the true observed spectrum is
synchrotron+blackbody and is fit with a Band+blackbody photon
model. We confirm the original conclusion of
\citet{Crider:1998,Preece:1998} that neither SCS or FCS alone can
explain the entire catalogs. Moreover, we find that if the true
observed spectrum is SCS then the ``line-of-death'' should actually be
at $\alpha_{\rm LOD}\simeq-0.8$ rather than the originally stated -2/3
owing to the fact that synchrotron asymptotically approaches a power
law shape and continuously curves below its $\vFv$ peak. This causes a
fit with the Band function to recover an $\alpha$ that is dependent on
the location of the $\vFv$ peak with respect to the GBM low energy
bandpass. Furthermore, we conclude that it is difficult to recover the
parameter distributions of the GBM spectral catalog by adding on a
blackbody to synchrotron emission if past fits to observations
represent an actual sample of the typical blackbody parameters. If GBM
observations typically contain a blackbody with $kT\simeq 30$ keV and
$R\simeq 0.1$ then values of $\alpha>-2/3$ would not be found when
fitting these spectra with a Band function alone. Also, much harder
values of Band's $\beta$ would be observed for the typically found
parameters of the blackbody. This is due to the broad total curvature
introduced by the combined synchrotron+blackbody spectrum which is too
broad for the narrower Band function to fit.

To explain the hardest $\alpha$-values in the catalog, we need to
observe blackbodies with a higher value of $R$ and the blackbody would
dominate the spectrum unlike what is observed. This finding indicates
that while fits to these GBM spectra with Band+blackbody are to be
statistically better descriptions of the data, the resulting Band
function is very different from a synchrotron function ($\alpha$ too
hard and the spectral width is too narrow.) We have checked the
preliminary GBM time-resolved spectral catalog \citep{Yu:2014b} for
spectra with hard $\alpha$ and $\beta$ which these simulations
indicate could contain a bright blackbody component. However, out of
$\sim 1800$ spectra, very few have both hard $\alpha$ and $\beta$ and
those that do contain no statistically significant blackbody.

The $\Ep$ recovered from the Band+blackbody fits to the multicomponent
simulations is not always accurate and can differ from the simulated
true value. This, combined with the fact that the fitted flux and $kT$
of the blackbody in these synthetic data are not always accurate means
that using the fitted Band function $\Ep$ in multicomponent fits to
infer properties about the GRB or to examine flux-luminosity relations
should be done with caution. The Band function is simply too flexible
and the free parameters work together to fit the curvature of the data
in the best way possible without regard for physics.

There are other forms of synchrotron emission and processes that can
result in different spectral shapes. For example, Klein-Nishina losses
can significantly alter the low-energy spectrum of synchrotron
emission \citep{Daigne:2011}. Additionally, \citet{Uhm:2014} have
shown that altering the magnetic field structure along the radial
direction of the outflow can also modify the low-energy
slope. However, it is not clear if these processes alter the spectral
curvature of synchrotron resulting in a narrower $\vFv$ peak more
consistent with the Band function. If that is the case, then it is
possible that the combination of this narrower synchrotron emission
produced in Poynting flux jets \citep{Giannios:2004,Zhang:2011} and a
blackbody could explain the GBM catalog; therefore, such emission
mechanism should be tested in a similar way as what is done here
\citep[see however][where a problem with Poynting flux jets and
photospheric emission is discussed]{Begue:2014}.

One should also note that the few studies that have attempted to
numerically simulate spectra composed of synchrotron emission and a
photospheric blackbody. For example, \citet{Hascoet:2013,Gao:2014} use
the \emph{Band function} with an $\alpha=-3/2$ as a proxy for the
actual synchrotron emission. This artificially imposes a narrower
curvature on the simulated spectra and guarantees that the spectra
will mimic the shape observed in the data. It will therefore be
difficult to use these simulations to assess the physical validity of
Band+blackbody fits to observed data. However, when these simulations
advance to the point that both the thermal and non-thermal components
are realistic physical representations of the theorized emission, a
similar assessment to what is done here can proceed.

Studies where a physical synchrotron photon model is used to replace
the Band function in spectral fits have shown that it is possible to
fit some GRB spectra with synchrotron or synchrotron+blackbody
\citep{Burgess:2014a,Yu:2014}. However, these works find that only SCS
can fit the data accurately due mainly to the \emph{curvature} of the
data around the $\vFv$ peak. However, calling these spectra SCS could
be misleading. \citet{Beniamini:2013} point out that when
$\gamma_{\min}$ and $\gamma_{\rm cool}$ are close to each other, the
electron distribution is in a marginally fast-cooling state but still
mimics the shape of SCS. Additionally, the narrowness of the $\vFv$
peak in the data can be fit with thermal emissivities related to
sub-photospheric dissipation \citep{Iyyani:2014}. All of these
findings indicate that spectral curvature must be considered when
trying to infer physics from fits to GRB spectra. There is a tendency
in theoretical modeling to aim for a single value of $\alpha$ as a
mark of success in explaining the emission process in GRBs. Obviously,
as we have shown, many factors such as curvature, detector bandpass,
and the limited shape of the Band function should be considered as
well as the peak of the $\alpha$ distribution when assessing the
predictive power of a model. \citet{Beloborodov:2013} points out that
synchrotron faces a problem of more than just the LOD. One must
consider the narrowness of the $\vFv$ peak and the clustering of $\Ep$
which is not easily reconciled with the current knowledge of electron
acceleration processes and astrophysical magnetic fields. In fact,
\citet{Axelsson:2014} study the width of spectral curvature in the GBM
catalogs and find that nearly half of the spectra are far too narrow
to be explained by synchrotron emission. It is possible that this
corresponds to our finding that only half of $\alpha$ distribution can
be explained by synchrotron or synchrotron+blackbody.

In conclusion, optically-thin synchrotron emission with or without a
blackbody accounting for emission from a non-dissipative photosphere
is insufficient to explain the \emph{entire} GBM spectral
catalog. Both scenarios can account for little more than half the
observed spectra. This implies that at least a large fraction of the
catalog GRBs have another origin such as emission from the photosphere
including subphotospheric-dissipation
\citep{Rees:2005,Peer:2005,Beloborodov:2010} and structured jets
\citet{Goodman:1986,Lundman:2014}. 

\section*{Acknowledgments}
We thank Andrei Beloborodov, Christoffer Lundman, and Bing Zhang for
interesting discussions on GRB emission mechanisms and how they affect
the spectral curvature of the data.  

\bibliographystyle{mn2e}
\bibliography{bib}

\label{lastpage}

\end{document}